\documentclass[a4paper,11pt]{article}
\pdfoutput=1 

\usepackage{jcappub} 
                    
\usepackage{xcolor}

\usepackage[T1]{fontenc} 
\usepackage{cancel}
\usepackage[normalem]{ulem}
\title{\boldmath Inflation in Symmergent Metric-Palatini Gravity}

\author[a]{Nilay Bostan,}
\author[b]{Canan Karahan,}
\author[c]{Ozan Sarg{\i}n}

\affiliation[a]{Proton Accelerator Facility, Turkish Energy Nuclear and Mineral Research Agency, Nuclear Energy Research Institute, 06980, Ankara, Türkiye}
\affiliation[b]{National Defence University, Turkish Naval Academy, Department of Basic Sciences, 34942 Tuzla, {\.I}stanbul, Türkiye}
\affiliation[c]{Sabanc{\i} University, Faculty of Engineering and Natural Sciences, 34956 Tuzla, {\.I}stanbul, Türkiye}
\emailAdd{nilay.bostan@tenmak.gov.tr}
\emailAdd{ckarahan@dho.edu.tr}
\emailAdd{ozan.sargin@sabanciuniv.edu}

\abstract{In this paper, we study the cosmological inflation phenomenon in symmergent gravity theory. Symmergent gravity is a novel framework which merges gravity and the standard model (SM) so that the gravity emerges from the matter loops and restores the broken gauge symmetries along the way. Symmergent gravity is capable of inducing the gravitational constant $G$ and the quadratic curvature coefficient $c_O$ from the loop corrections of the matter sector in a flat space-time. In the event that all the matter fields, including the beyond the standard model (BSM) sector, are mass degenerate, the vacuum energy can be expressed in terms of $G$ and  $c_O$. The parameter which measures the deviation from the mass degeneracy is dubbed $\hat{\alpha}$. The parameters, $c_O$ and $\hat{\alpha}$, of symmergent gravity convey the information about the fermion and boson balance in the matter (SM+BSM) sector in number and in mass, respectively. In our analysis, we have investigated the space of the symmergent parameters $c_O$ and $\hat{\alpha}$  wherein they produce results that comply with the inflationary observables $n_s$, $r$, and $\mathrm{d}n_s/\mathrm{d}\ln k$.   

We have shown that the vacuum energy together with the quadratic curvature term arising in the symmergent gravity prescription are capable of inflating the universe provided that the quadratic curvature coefficient $c_O$ is negative (which corresponds to fermion dominance in number in the matter sector) and the deviation from the mass degeneracy in the matter sector is minute for both boson mass dominance and fermion mass dominance cases.  }


\begin{document}
\maketitle
\flushbottom

\section{Introduction}
\label{sec:intro}

The Standard Model (SM), which elegantly expresses the fundamental particles that make up the visible matter in the universe and the interactions between these particles, is the result of nearly a hundred years of work, both theoretically and experimentally. This model has been completed in terms of its predictions of particle spectrum with the discovery of the Higgs boson at the Large Hadron Collider (LHC) \cite{higgs}. Regardless of all  achievements of SM at Fermi energy scales, there still exists many problems such as dark matter \cite{Zwicky:1933gu,Freese:2017idy}, neutrino masses \cite{solar,atmospheric}, incorporation of gravity into SM, etc., which are to be undertaken. All these problems demonstrate that there should be new physics beyond the SM. Furthermore, lack of any new physics sign at experiments tempts one to think about SM to be valid up to the Planck scale, which is the scale of gravity. However, SM is an effective quantum field theory (QFT) in flat space-time, which possess hypersensitivity to the ultraviolet (UV) scale. To put it more explicitly, gauge and scalar boson masses, and also vacuum energy gets huge loop corrections  proportional to quadratic and quartic power of UV cutoff scale $\Lambda$. Getting these UV corrections breaks the gauge symmetries of QFT explicitly. To restore gauge symmetries, there is a novel  approach called symmergent gravity, which eliminates  all UV sensitivies of QFT via emergence of gravity 
 \cite{demir2, demir3, bizimki}.

Symmergent gravity may be considered as a special form of affine $f(R)$ gravity \cite{De Felice2010}, which incorporates a quadratic curvature term, whose parameters originate from loop factors of a flat space time QFT. These loop factors bear the footprints of the boson-fermion balance both in number and mass in the new physics sector. Studying cosmological implications of symmergent gravity is essential for obtaining crucial information about the nature of new physics. Recently, black hole studies have given critical signatures of symmergent gravity \cite{Cimdiker:2021cpz,Rayimbaev:2022hca,Pantig:2022qak,Gogoi:2023fow}. Cosmological inflation is pivotal  to constrain the gravitational theories. Therefore, it is also expected to be suitable to investigate symmergent gravity in the inflationary setting. There is a vast literature on inflationary cosmology  \cite{Martin:2013tda}. Amongst the diluge of inflationary models the single field models are the predominant ones. A large number of these single field models calculate the inflationary predictions by considering a non-minimal coupling between the inflaton and the curvature (see \cite{Bezrukov:2010jz, Bostan:2018evz, Bezrukov:2007ep,Tenkanen:2017jih,Bauer:2008zj,Rasanen:2017ivk,Jinno:2019und,Rubio:2019ypq,Bostan:2022swq,Enckell:2018kkc, Bostan:2019wsd, Tenkanen:2019jiq,Jarv:2020qqm, nilay,Linde:2011nh} and references therein).    

In this work, we study cosmological inflation in symmergent gravity framework with a constant potential. The constant potential used in this work is the vacuum energy of matter sector which is not power-law in the cutoff. Cosmological inflation in symmergent gravity without considering the effect of vacuum energy has been studied \cite{irfan} analytically via a comparison with the Starobinsky inflation. We demonstrate that, to be able to satisfy the requirements of inflation in accordance with symmergent gravity, fermions should outnumber the bosons in the matter sector (SM+BSM) regardless of whether the bosons or the fermions have the mass dominance. 

The paper is organized as follows: in section \ref{Sec.2}, the symmergent gravity and the related cosmological parameters of inflation, such as slow-roll parameters with respect to canonical field ($\sigma$), are introduced. Then, in section \ref{analysis}, the results of our inflationary analysis are given. Finally, the discussion of our results has been undertaken in section \ref{conc}.


\section{Model}\label{Sec.2}
\subsection{Symmergent Gravity}

Symmergent gravity is a new type of emergent gravity theory which prioritizes  restoring the broken gauge symmetries  \cite{demir2, demir3, bizimki}. In this manner, it differs from other emergent gravity theories, such as Sakharov's setup. In general, a UV momentum cutoff $\Lambda$ on QFTs breaks explicitly all the gauge symmetries. This is because, a QFT of scalars $\phi$ and gauge bosons $V_\mu$ with the action $S[\eta, \phi, V_\mu]$  gets the one loop correction
\begin{eqnarray}
\label{QFT-action}
\delta S[\eta,\phi,V_{\mu}]=\int d^4x \sqrt{-\eta}\left(-V_0-c_O\Lambda^4-\sum_{i}c_{m_i} {m_i}^2\Lambda^2-c_{\phi}\Lambda^2 \phi^{\dagger}\phi+c_V\Lambda^2 V_{\mu}V^{\mu}\right),
\end{eqnarray}
where $V_0$  is the vacuum energy that does not have $\Lambda$, $c_O$ is the loop factor describing the quartic vacuum energy correction, $c_\phi$ is the loop factor of quadratic scalar mass correction, $c_V$ is the loop factor of loop-induced gauge boson mass, and $c_m$ is the loop factor of quadratic vacuum energy correction. Here, $\eta_{\mu\nu}$ is the flat metric with the signature  $(+,-,-,-)$.

Typically, gauge fields acquire a mass term like $c_V \Lambda^2 V_\mu V^\nu$ as in  (\ref{QFT-action}). This gauge boson mass term is responsible for the explicit symmetry breaking because the UV momentum cutoff $\Lambda$ is not a Casimir invariant of the Poincare group. Sakharov associates $M_{Pl}$ with the momentum cutoff $\Lambda$, and this results in a Planckian sized cosmological constant and Planckian sized scalar masses. It also falls short in ameliorating the broken gauge symmetry. Symmergent gravity, on the other hand, approaches the problem from a different angle by claiming that this explicit gauge boson mass term can be alleviated by using the famous Higgs mechanism. Here, it is important to realize that the UV cutoff  $\Lambda$ is a Poincare (translation) breaking scale and the corresponding Higgs field must also be a Poincare breaking one. This leads one to discern that the sought-for Higgs field is the affine curvature \cite{demir2, demir3} defined via
\begin{eqnarray}
{\mathbb{R}}_{\mu\nu}(\Gamma) = \partial_\lambda \Gamma^\lambda_{\mu\nu} -\partial_\nu \Gamma^\lambda_{\mu\lambda} + \Gamma^\rho_{\lambda\rho} \Gamma^\lambda_{\mu\nu}  - \Gamma^\lambda_{\rho\nu}\Gamma^\rho_{\mu\lambda},
\label{affine-curv}
\end{eqnarray}
as the curvature of the affine connection  $\Gamma^\lambda_{\mu\nu}$, which is completely independent of the Levi-Civita connection ${}^g\Gamma^\lambda_{\mu\nu}$ of the metric tensor $g_{\mu\nu}$. 

The prescription of the symmergent gravity framework in order to remedy the gauge symmetry breaking problem  can be described as follows. First, one takes the effective QFT in (\ref{QFT-action}) to a curved space time with metric $g_{\mu\nu}$. Second, one carries out a  map in the gauge boson mass term 
\begin{equation}
    c_V \Lambda^2 V_\mu V^\mu \rightarrow c_V V_\mu \left(\Lambda^2 g^{\mu\nu} - R^{\mu\nu}(g)\right)  V_\nu \;.
\end{equation}
In doing this map one uses the notion that the Ricci curvature of the curved metric can show up only in the gauge sector via the covariant derivatives. Third, one promotes UV cutoff  $\Lambda$ to affine curvature just like the promotion of vector boson masses to the Higgs field:
\begin{equation}
    \Lambda^2 g^{\mu\nu} \rightarrow {\mathbb{R}}^{\mu\nu}(\Gamma) \;.
\end{equation}
As a result, the effective action in (\ref{QFT-action}) leads to the metric-Palatini action \cite{biz-beyhan,affine3} 
\begin{eqnarray}
S_{P}[g, {\mathbb{R}}, R]  &=& \int d^4x \sqrt{-g} \Bigg\{\!\!-V_0
-\frac{1}{16\pi G}{\mathbb{R}}(g,\Gamma)   - \frac{c_O}{16} \left({\mathbb{R}}(g,\Gamma)\right)^2  -\frac{c_\phi}{4} \phi^{\dagger}\phi  \; {\mathbb{R}}(g,\Gamma) \nonumber\\&&\qquad\qquad\quad\;\;+\, c_V {\rm tr}\left[V^{\mu}\!\left({\mathbb{R}}_{\mu\nu}(\Gamma)- R_{\mu\nu}({}^g\Gamma)\right)\!V^{\nu}\right]\!\Bigg\},
\label{Sp-curvp}
\end{eqnarray}
in which ${\mathbb{R}}(g,\Gamma)\equiv g^{\mu\nu}{\mathbb{R}}_{\mu\nu}(\Gamma)$ is the affine curvature scalar, and $\phi$ is a real scalar field.

Here, it is time to elaborate on the Newton's constant $G$, loop factors $c_O$, $c_V$ and $c_\phi$ that depend on the details of the underlying QFT. 
The Newtons's constant $G$, arising from the quadratic vacuum energy correction term in (\ref{QFT-action}), is expressed in terms of mass-squared matrix $\mathcal{M}^2$  of all fields as 
\begin{eqnarray}
G^{-1}=\frac{1}{8\pi}str[\mathcal{M}^2],
\end{eqnarray}
in which $str[\mathcal{M}^2]=\sum_i\left(-1\right)^{2s_i}\left(2s_i +1\right)tr[\mathcal{M}^2]_{s_i}$ ( $str[...]$ stands for super-trace) where $s_i$ is the spin of the QFT fields $\psi_i(s_i=0,1/2,...)$.

The quadratic curvature coefficient $c_O$, which is induced as the loop factor of the quartic vacuum energy correction metamorphoses into the quadratic curvature coefficient in symmergent gravity and has the value at one loop  \cite{bizimki}
\begin{eqnarray}
c_O=\frac{n_b-n_f}{64 \pi^2},
\label{cO}
\end{eqnarray}
which is universal in that here $n_b (n_f)$ is the number of bosons (fermions) having masses from zero way up to the gravitational scale.

The vacuum energy density $V_0$ in Eq.(\ref{Sp-curvp}) is defined as $str[\mathcal{M}^4]/64\pi^2$, and it is equivalent to
\begin{eqnarray}
V_0=\frac{1}{64\pi^2}\left(\sum_{b}m_b^4-\sum_{f}m_f^4\right),
\label{V_0}
\end{eqnarray}
where $m_b$ ($m_f$) stands for the boson (fermion) mass. 

In the limit where  all the boson and fermion masses are set to a common mass $M_0$ (mass-degenerate limit), which can be considered as the characteristic scale of the QFT, vacuum energy may be expressed in terms of Newton's constant and the loop factor $c_O$ as follows
\begin{eqnarray}
V_0=\frac{M_0^4}{64\pi^2}\left(n_b-n_f\right)=\frac{1}{2(8\pi G)^2c_O}.
\label{V_0_degeneracy}
\end{eqnarray}
If one wants to consider a QFT of characteristic scale $M_0$ but whose  mass spectrum is not necessarily mass-degenerate, a more realistic expression for vacuum energy may be written as 

\begin{eqnarray}
V_0=\frac{1-\hat{\alpha}}{(8\pi G)^2c_O},
\label{V_0_real}
\end{eqnarray}
in which $\hat{\alpha}$ is a parameter expressing the deviation from mass-degeneracy. In this expression, $\hat{\alpha}>1$ ($\hat{\alpha}<1$) corresponds to boson (fermion) mass dominance.

In order to make the equations compact, we will keep $V_0$ in closed form until the inflationary analysis part in section \ref{analysis}.

 The affine connection in the metric-Palatini action (\ref{Sp-curvp}) can be eliminated dynamically by solving its equation of motion (namely $\delta_\Gamma S_{P}[g, {\mathbb{R}}, R] = 0$)

\begin{eqnarray}
\label{gamma-eom}
{}^{\Gamma}\nabla_{\lambda} (\sqrt{-g}\;{\Tilde{\mathbb{D}}_{\mu\nu}}) = 0,
\end{eqnarray}
such that ${}^{\Gamma}\nabla_{\lambda}$ is the covariant derivative of the affine connection $\Gamma^\lambda_{\mu\nu}$, and 

\begin{eqnarray}
\label{q-tensor}
{\Tilde{\mathbb{D}}_{\mu\nu}} = \left(\frac{1}{2} M_{Pl}^{2} +   \frac{c_\phi}{4} \phi^2 + \frac{c_O}{8} {\mathbb{R}}(g,\Gamma)\right) g_{\mu\nu} - c_{V} {\mbox{tr}}\left[V_{\mu}V_{\nu}\right]
\end{eqnarray}
is the disformal metric of tensor fields, including the affine curvature ${\mathbb{R}}(\Gamma)$ itself.  The motion equation (\ref{gamma-eom}) 
can be recast into
\begin{eqnarray}
\label{gamma-eom1}
{}^{\Gamma}\nabla_{\lambda} (\sqrt{-\mathbb{D}}\;{\mathbb{D}_{\mu\nu}}) = 0,
\end{eqnarray}
where 
\begin{eqnarray}
\label{q-tensor1}
{\mathbb{D}_{\mu\nu}} = \left(\frac{g}{\Tilde{\mathbb{D}}}\right)^{1/6}{\Tilde{\mathbb{D}}_{\mu\nu}} \: .
\end{eqnarray}
This new form of the motion equation (\ref{gamma-eom1}) implies that ${\mathbb{D}}_{\mu\nu}$ is covariantly-constant with respect to $\Gamma^\lambda_{\mu\nu}$, and this constancy leads to the exact solution 
\begin{eqnarray}
\Gamma^\lambda_{\mu\nu} &=& \frac{1}{2} ({\mathbb{D}}^{-1})^{\lambda\rho} \left( \partial_\mu {\mathbb{D}}_{\nu\rho} + \partial_\nu {\mathbb{D}}_{\rho\mu} - \partial_\rho {\mathbb{D}}_{\mu\nu}\right)\nonumber\\
&=& {}^g\Gamma^\lambda_{\mu\nu} + \frac{1}{2} ({\mathbb{D}}^{-1})^{\lambda\rho} \left( \nabla_\mu {\mathbb{D}}_{\nu\rho} + \nabla_\nu {\mathbb{D}}_{\rho\mu} - \nabla_\rho {\mathbb{D}}_{\mu\nu}\right),
\label{aC}
\end{eqnarray}
in which, needless to say, ${}^g\Gamma^\lambda_{\mu\nu}$ is the Levi-Civita connection of the curved metric $g_{\mu\nu}$. 

The Planck scale is the largest scale, and therefore it is legitimate to expand the affine connection (\ref{aC}) and obtain the curvature  in terms of $M_{Pl}$. In order to accomplish this, we first define
\begin{eqnarray}
{\Tilde{\mathbb{D}}_{\mu\nu}} = \frac{1}{2} M_{Pl}^{2} \; g_{\mu\nu} + \Delta_{\mu \nu},
\end{eqnarray}
where
\begin{eqnarray}
\Delta_{\mu \nu} = \left( \frac{c_\phi}{4} \phi^2 + \frac{c_O}{8} {\mathbb{R}}(g,\Gamma)\right) g_{\mu\nu} - c_{V} {\mbox{tr}}\left[V_{\mu}V_{\nu}\right] \; .
\label{deltamunu}
\end{eqnarray}
In powers of $M_{Pl}$, the affine connection takes the following form
\begin{eqnarray}
\Gamma^{\lambda}_{\mu\nu}&=&{}^{g}\Gamma^{\lambda}_{\mu\nu} + \frac{1}{M_{Pl}^2} \left( \nabla_\mu {\overline{\mathbb{D}}^\lambda_\nu }+ \nabla_\nu {\overline{\mathbb{D}}^\lambda_\mu} - \nabla^\lambda {\overline{\mathbb{D}}_{\mu\nu}}\right) + {\mathcal{O}}\left(M_{Pl}^{-4}\right),
\label{expand-conn}
\end{eqnarray}
in which
\begin{eqnarray}
{\overline{\mathbb{D}}_{\mu\nu}} = \Delta_{\mu \nu} - \frac{1}{6} \Delta^{\alpha}_{\; \alpha} \; g_{\mu\nu} \;.
\end{eqnarray}
The curvature is obtained up to ${\mathcal{O}}\left(M_{Pl}^{-2}\right)$ as follows
\begin{equation}\label{expand-curv}
{\mathbb{R}}_{\mu\nu}(\Gamma) = R_{\mu\nu}({}^{g}\Gamma) + \frac{1}{M_{Pl}^2}\left(\nabla^{\alpha} \nabla_{\mu} {\overline{\mathbb{D}}}_{\alpha\nu} + \nabla^{\alpha} \nabla_{\nu} {\overline{\mathbb{D}}}_{\alpha\mu} - \Box {\overline{\mathbb{D}}}_{\mu\nu} - \nabla_{\mu} \nabla_{\nu} {\overline{\mathbb{D}}}_{\alpha}^{\alpha}\right)  +  {\mathcal{O}}\left(M_{Pl}^{-4}\right),
\end{equation}
so that both $\Gamma^{\lambda}_{\mu\nu}$ and ${\mathbb{R}}_{\mu\nu}(\Gamma)$ contain  pure derivative terms  at the next-to-leading ${\mathcal{O}}\left(M_{Pl}^{-2}\right)$ order \cite{demir2,demir3}. 

The expansion in (\ref{expand-conn}) ensures that the affine connection $\Gamma^{\lambda}_{\mu\nu}$ is solved algebraically order by order in $1/M_{Pl}^{2}$  despite the fact that its motion equation (\ref{gamma-eom}) involves its own curvature ${\mathbb{R}}_{\mu\nu}(\Gamma)$ through $\overline{{\mathbb{D}}}_{\mu\nu}$   \cite{demir3}. The expansion (\ref{expand-curv}), on the other hand, ensures that the affine curvature  ${\mathbb{R}}_{\mu\nu}(\Gamma)$ is equal to the metrical curvature $R_{\mu\nu}({}^g\Gamma)$ up to a doubly-Planck suppressed remainder. In essence, what happened is that the affine dynamics took the affine curvature ${\mathbb{R}}$ from its UV value $\Lambda_\wp^2$   to its IR value $R$ in eq. (\ref{expand-curv}). Indeed, in the sense of holography \cite{holog1}, the metrical curvature $R$ sets the IR scale \cite{holog2} above which QFTs hold as flat spacetime constructs \cite{wald1,cgr2}. 

Under the affine curvature in eq. (\ref{expand-curv}), the metric-Palatini action $S_{P}[g, {\mathbb{R}}, R]$  in eq. (\ref{Sp-curvp}) 
reduces to the GR plus quadratic curvature term 
\begin{eqnarray}\label{reduce-nongauge}
S_{P}[g, {\mathbb{R}}, R] = \int\!\! d^4x \sqrt{-g} \Bigg\{\!\!\! 
&-&V_0-\frac{M_{Pl}^2}{2}R   - \frac{c_O}{16} R^2  -\frac{c_\phi}{4} \phi^2 R \\ \nonumber
&-& \frac{2\Delta^{\mu\nu}}{M_{Pl}^2} \! \left(\nabla^{\alpha} \nabla_{\nu} {\Delta}_{\alpha\mu} -\frac{1}{3} \nabla_{\nu} \nabla_{\mu} {\Delta}^{\kappa}_{\;\kappa} - \frac{1}{2} \Box {\Delta}_{\mu\nu} +\frac{1}{12} g_{\mu \nu} \Box  {\Delta}^{\kappa}_{\;\kappa}\right) \\ \nonumber
&+& {\mathcal{O}}\!\left(M_{Pl}^{-4}\right)\!\!\Bigg\},
\end{eqnarray}
in which $R=g^{\mu\nu}R_{\mu\nu}({}^g\Gamma)$ is the usual curvature scalar in the GR and ${\Delta}_{\mu\nu}$ is defined in eq. (\ref{deltamunu}).

One notices that the gauge field contribution in eq. (\ref{Sp-curvp}) vanishes at the leading order. The remainder involves derivatives  of the long-wavelength fields $\phi$ and $V_\mu$, produces thus  no mass terms for these fields, and remains small for all practical purposes.  

In a cosmological spacetime (like Friedmann–Robertson–Walker (FRW)), the vector fields $V_\mu$ cannot form a background (excepting SU(2) gauge fields). In such a case, the action (\ref{reduce-nongauge}) takes the form:
\begin{eqnarray}\label{reduce-nongauge2}
S_{cos}=\int d^4x \sqrt{-g} \Bigg\{&-&V_0-
\frac{M_{Pl}^2}{2}R   - \frac{c_O}{16} R^2  -\frac{c_\phi}{4} \phi^2 R +\frac{1}{2} \partial_\mu\phi \partial^\mu \phi  -V(\phi)\nonumber\\
&+&\frac{2}{M_{Pl}^2} \! \left(\frac{c_\phi}{4} \phi^2 + \frac{c_O}{8} R\right) \Box \left(\frac{c_\phi}{4} \phi^2 + \frac{c_O}{8} R\right) + {\mathcal{O}}\!\left(M_{Pl}^{-4}\right)\Bigg\},
\end{eqnarray}
after restoring the kinetic and potential energy densities of the scalar field.

\subsection{Cosmological Inflation}
At first glance, the cosmological action (\ref{reduce-nongauge2}) seems to be too complicated to investigate the cosmological inflation because it incorporates the non-minimally coupled fourth-order and sixth-order gravity terms.  This paper constitutes the first step towards the investigation into the full action (\ref{reduce-nongauge2}). Therefore, the rest of the paper is devoted to the case, where we ignore the $\phi$ sector and  $M_{Pl}^2$ suppressed terms but keep the vacuum energy density term $V_0$. This choice reduces (\ref{reduce-nongauge2}) to the simpler form 
\begin{eqnarray}
S_{cos}=\int d^4x \sqrt{-g} \Bigg\{&-&
\frac{M_{Pl}^2}{2}R   - \frac{c_O}{16} R^2  -V_0\Bigg\}.
\label{reduce-nongauge3}
\end{eqnarray}

This fourth-order gravitational action can be cast into a non-minimally coupled auxiliary field action in the Jordan frame via the scalar-tensor theory and f(R) gravity equivalance. Doing this, (\ref{reduce-nongauge3}) takes the form 

\begin{eqnarray}
S_{J}=\int d^4x \sqrt{-g} \Bigg\{&-&
\frac{M_{Pl}^2}{2}\left(1+\frac{c_O}{4M_{Pl}^2}\chi^2\right)R+\frac{c_O}{16}\chi^4  -V_0\Bigg\},
\label{reduce-nongauge4}
\end{eqnarray}
 where $\chi$ is the auxiliary scalar. In this action, the total potential in Jordan frame reads
 \begin{eqnarray}
 V_J=V_0-\frac{c_O}{16}\chi^4.
 \label{Jrd_pot}
 \end{eqnarray} 
By performing the conformal transformation
\begin{eqnarray}
\bar{g}_{\mu\nu}=F(\chi){g}_{\mu\nu},
\label{conf_trans_1}
\end{eqnarray}
in which 
\begin{eqnarray}
F(\chi)=\left(1+\frac{c_O}{4M_{Pl}^2}\chi^2\right),
\end{eqnarray}
the Einstein frame action is obtained as follows
\begin{eqnarray}
S_{E}=\int d^4x \sqrt{-\bar{g}} \Bigg\{&-&
\frac{M_{Pl}^2}{2}R+\frac{1}{2}\partial_{\mu}\sigma \, \partial^{\mu}\sigma  -V_E(\sigma)\Bigg\},
\label{reduce-nongauge5}
\end{eqnarray}
where the potential in Einstein frame in terms of auxiliary field $\chi$ is defined as 
\begin{eqnarray}
V_E(\chi)=\frac{V_J(\chi)}{F(\chi)^2} = \frac{V_0-\frac{c_O}{16}\chi^4}{\left(1+\frac{c_O}{4M_{Pl}^2}\chi^2\right)^2},
\end{eqnarray}
and the auxiliary field is dressed into the canonically normalized field 
\begin{eqnarray}
 \sigma =  M_{Pl} \:  \sqrt{\frac{3}{2}} \; \ln F(\chi) \;.
\end{eqnarray}
In terms of canonical field $\sigma$, the potential in Einstein frame reads out
\begin{eqnarray}
    V_E(\sigma)=\frac{V_0-\Big[\frac{M_{Pl}^4}{c_O} \left(e^{\sqrt{\frac{2}{3}}\frac{\sigma}{M_{Pl}}}-1\right)^2\Big]}{e^{2\sqrt{\frac{2}{3}}\frac{\sigma}{M_{Pl}}}}.
    \label{canonicpot}
\end{eqnarray}


\subsection{Inflationary observables}

By using slow-roll parameters, inflationary predictions can be defined \cite{Lyth:2009zz}. The slow-roll parameters in terms of the canonical field $\sigma$ have the forms 
\begin{equation}\label{slowroll1} 
\epsilon =\frac{M^2_{Pl}}{2}\left( \frac{V_{E, \sigma} }{V_E}\right) ^{2}\,, \quad
\eta = M^2_{Pl}\frac{V_{E, \sigma\sigma} }{V_E}  \,, \quad
\zeta ^{2} = M^4_{Pl} \frac{V_{E, \sigma} V_{E, \sigma \sigma\sigma} }{V_E^{2}}\,,
\end{equation}
wherein the subscripts $\sigma$'s imply derivatives.

In the slow-roll approximation, inflationary observables, $n_s, r, \alpha=\mathrm{d}n_s/\mathrm{d}\ln k$, can be described as
\begin{eqnarray}\label{nsralpha1}
n_s = 1 - 6 \epsilon + 2 \eta \,,\quad
r = 16 \epsilon, \nonumber\\
\alpha=\frac{\mathrm{d}n_s}{\mathrm{d}\ln k} = 16 \epsilon \eta - 24 \epsilon^2 - 2 \zeta^2\,.
\end{eqnarray}

Here, $n_s$ is the spectral index, $r$, the tensor-to-scalar ratio, and $\alpha=\mathrm{d}n_s/\mathrm{d}\ln k$ is the running of the spectral index. In addition, the number of e-folds, in the slow-roll approximation, can be written in the following form
\begin{equation} \label{efold1}
N_*=\frac{1}{M^2_{Pl}}\int^{\sigma_*}_{\sigma_e}\frac{V_E \ \rm{d}\sigma}{V_{E, \sigma}}\, , \end{equation}
here, the subscript ``$_*$'' implies the quantities at which the scale corresponding to $k_*$ exited the horizon. Moreover, $\sigma_e$ describes inflaton value at which the inflation ends. One can easily find $\sigma_e$ by using $\epsilon(\sigma_e) = 1$. The number of e-folds should be in the range between $\approx$ 55-60. On the other hand, it is pivotal to use the exact value of $N_*$ which depends on universe evolution. Assuming the standard thermal history after inflationary era, $N_*$ can be found, and inflationary predictions can be computed accordingly. With this assumption, $N_*$ takes the form \cite{Liddle:2003as}
\begin{eqnarray} \label{efolds}
N_*\approx64.7+\frac12\ln\frac{\rho_*}{M^4_{Pl}}-\frac{1}{3(1+\omega_r)}\ln\frac{\rho_e}{M^4_{Pl}} +\Big(\frac{1}{3(1+\omega_r)}-\frac14\Big)\ln\frac{\rho_r}{M^4_{Pl}}\,,
\end{eqnarray}
here, $\rho_{e}=(3/2)V_E(\sigma_{e})$ describes the energy density at the end of inflation and $\rho_r$ corresponds to the energy density at the end of reheating. Also, $\rho_{*} \approx V_E(\sigma_*)$ is the energy density at which the scales corresponding to $k_*$ that is exited the horizon. 

In addition, the curvature perturbation amplitude in terms of $\sigma$ can be defined in the following form
\begin{equation} \label{perturb1}
\Delta_\mathcal{R}^2=\frac{1}{12\pi^2 M^6_{Pl}}\frac{V_E^3}{V_{E, \sigma}^2},
\end{equation}
and this should be well-matched with the value of $\Delta_\mathcal{R}^2\approx 2.1\times10^{-9}$ from the Planck results \cite{Aghanim:2018eyx} for the pivot scale $k_* = 0.002$ Mpc$^{-1}$. In addition, $\omega_r$ is the equation of state parameter during reheating. Throughout this work, following the instant reheating assumption, we take $\omega_r=1/3$ while computing the inflationary predictions.

\begin{figure}\label{potsigma}
\centering
\includegraphics[width=.45\textwidth]{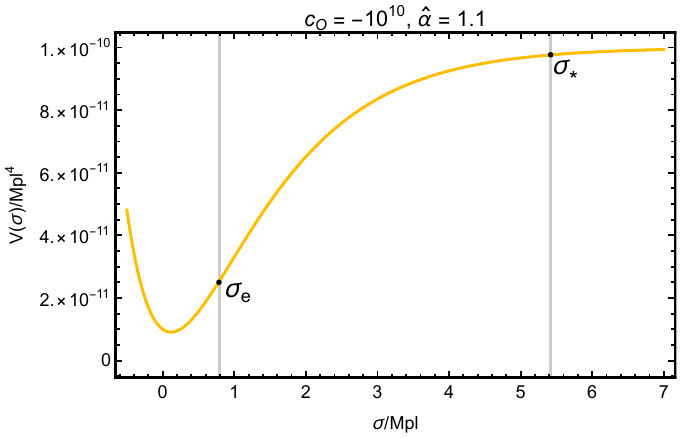}
\includegraphics[width=.45\textwidth]{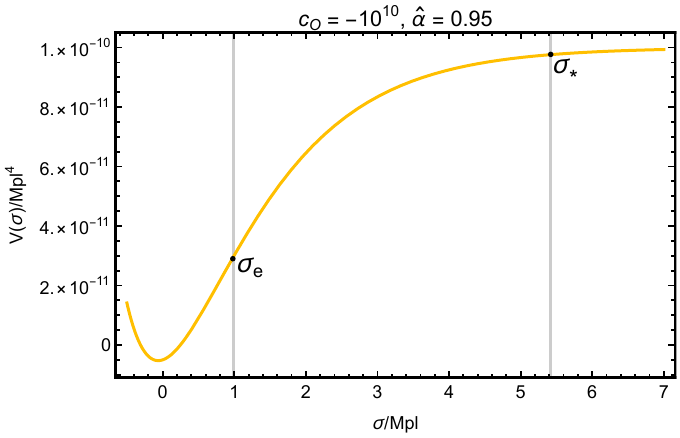}
\caption{\label{fig1} 
$\sigma/M_{Pl}$ vs. $V(\sigma)/M^4_{Pl}$ for $c_O = -10^{10}$, $\hat{\alpha} = 1.1 $ (left) and $\hat{\alpha} = 0.95$ (right).}
\end{figure}


\section{Analysis} 
\label{analysis}
In this section, we declare our results for the inflation potential  defined in  \eqref{canonicpot} together with the vacuum energy density $V_0$ given in \eqref{V_0_real}. 
This form of potential is in Einstein frame and in terms of canonical scalar field $\sigma$. It is important to note that here, for our numerical calculations in this section, we use the units where the reduced Planck scale $M_{Pl}= 1/ \sqrt {(8\pi G)} \approx 2.43 \times 10^{18} $ GeV.

The potential profile of the inflaton is depicted in figure \ref{fig1} for $c_O = -10^{10}$, $\hat{\alpha} = 1.1 $ (left panel) and $\hat{\alpha} = 0.95$ (right panel) cases. For $\hat{\alpha}=1.1$ $\rightarrow$ $\sigma_*\simeq 5.41676$ and $\sigma_e \simeq 0.799354$, as well as, for $\hat{\alpha}=0.95$ $\rightarrow$ $\sigma_*\simeq 5.41898$ and $\sigma_e \simeq 0.988536$. A comparison between the left and right panels of the figure \ref{fig1} asserts that, as $\hat{\alpha}$ increases, $\sigma_e$ decreases, even though $\sigma_*$ has a negligible change between two $\hat{\alpha}$ cases. This means that the duration of the inflationary epoch is greater for the $\hat{\alpha} > 1$, which corresponds to a matter sector in which bosons outweigh the fermions.
\begin{figure}\label{volarger1CL}
\centering
\includegraphics[width=.36\textwidth]{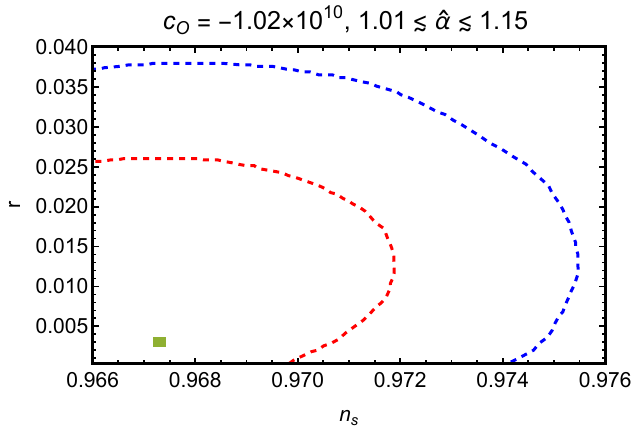} \qquad 
\includegraphics[width=0.47\textwidth]{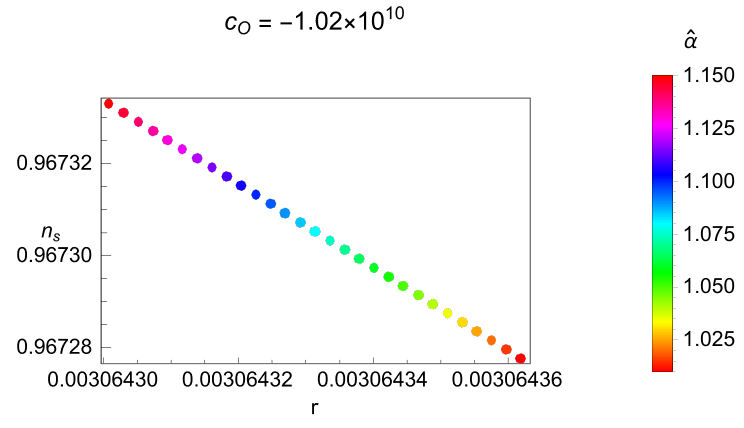}
\caption{\label{fig2} Inflationary predictions of the potential in \eqref{canonicpot}. The left panel shows $n_s-r$ values for $c_O = -1.02 \times 10^{10}$ and $1.01 \lesssim \hat{\alpha} \lesssim 1.15$ and blue(red) contours indicate the recent 95\%(68\%) CL given by BICEP/Keck \cite{BICEP:2021xfz}. Right panel shows $r-n_s$ values, where each dot corresponds to a different $\hat{\alpha}$ value (color coded) but to the same $c_O = -1.02 \times 10^{10}$.}
\end{figure}

Before going into the details of the numerical analysis, it is useful to give the inflationary predictions of the model from an analytical standpoint. The spectral tilt $n_s$ and the tensor-to-scalar ratio $r$ in our model have the asymptotic expressions in terms of the number of e-folds 
$N_*$ as 
\begin{eqnarray}\label{asymptotic_star}
    n_s\approx 1-\frac{2}{N_*}-\frac{9}{2 N_*^2}, \qquad r \approx \frac{12}{N_*^2},
\end{eqnarray}
whose numerical values for $N_*\simeq 60$ are $n_s\approx0.96541\overline{6}$ and $r\approx0.00\overline{3}$. These are exactly the same values that have been reported for the Starobinsky inflation and this makes sense because those asymptotic expressions in \eqref{asymptotic_star} are derived in the limit where $\sigma \gg M_{Pl}$ and our vacuum energy density term $V_0$ being proportional to the $M_{Pl}^4$ is ineffective in this limit.

The numerical analysis involves calculating the inflationary predictions $n_s, r$, and \newline
$\mathrm{d}n_s/\mathrm{d}\ln k$, for this potential and comparing our results with the latest constraints given by BICEP/Keck \cite{BICEP:2021xfz}. We display our results for two different cases: i) $\hat{\alpha} > 1$ and ii) $\hat{\alpha} < 1$. These show the effects of bosons and fermions dominance with regard to masses on the inflationary predictions. In our numerical results, instead of taking number of e-folds as a constant, such as $60$, we use $N_*$ which is given in eq. \eqref{efolds} supposing the standard thermal history after inflationary epoch.

Figures \ref{fig2} and \ref{fig4} display the inflationary predictions of the potential in \eqref{canonicpot}. In figure \ref{fig2}, the left panel presents $n_s-r$ values for $c_O = -1.02 \times 10^{10}$ and $1.01 \lesssim \hat{\alpha} \lesssim 1.15$ and the right panel presents $n_s-r$ values, where each dot corresponds to a different $\hat{\alpha}$ value (color coded) but to the same $c_O = -1.02 \times 10^{10}$. Similarly, in figure \ref{fig4}, the left panel shows $n_s-r$ values for $c_O = -1.035 \times 10^{10}$ and $0.13 \lesssim \hat{\alpha} \lesssim 0.97$ and right panel shows $n_s-r$ values, where each dot corresponds to a different $\hat{\alpha}$ value (color coded) but to the same $c_O = -1.035 \times 10^{10}$. In the left panels of both figures, blue(red) contours indicate the recent 95\%(68\%) CL given by BICEP/Keck \cite{BICEP:2021xfz}. 

For both $\hat{\alpha} > 1$ and $\hat{\alpha} < 1$ cases, we find $n_s\approx 0.967$ and $r\approx 0.003$, and these predictions remain inside the 68\% CL.  As is widely known, plateau type potentials have the property that the canonical field potential approaches to constant value for large field values. These are also known as the $R^2$ (Starobinsky) type inflation models \cite{Kehagias:2013mya} and for $N_*\simeq60$ e-folds they predict   $n_s\approx 0.966$ and $r\approx 0.0033$. These values of the predictions are also within the 68\% CL of the Planck results. Therefore, we can conclude that the predictions of our model is very close to the $R^2$ (Starobinsky) model, which are both very compatible with the observational data, residing inside the confidence regions of $1$-$\sigma$ in the latest constraints.
\begin{figure}
\label{volarger1nsr}
\centering
\includegraphics[width=0.88\textwidth]{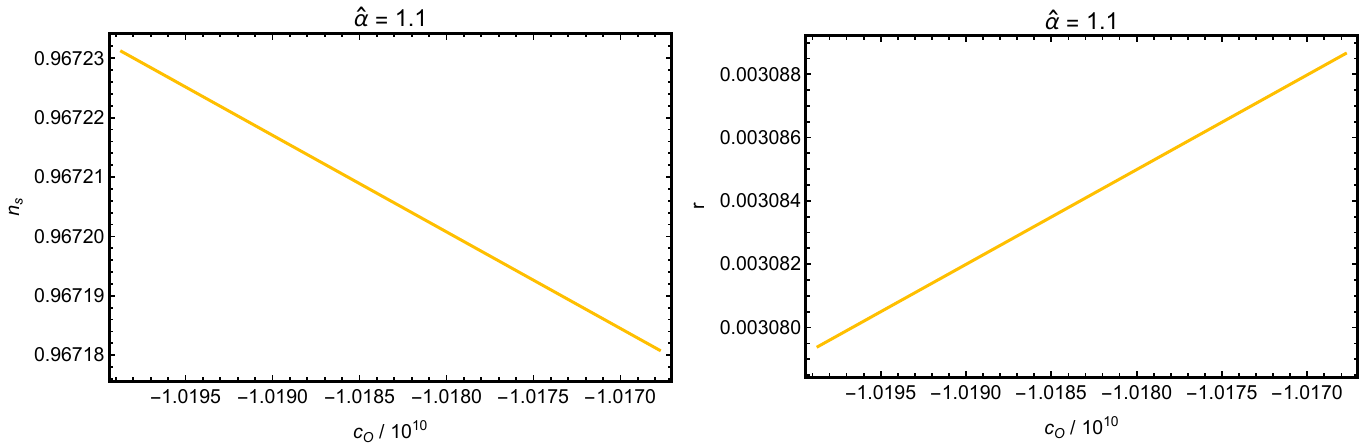}\\
\includegraphics[width=.4\textwidth]{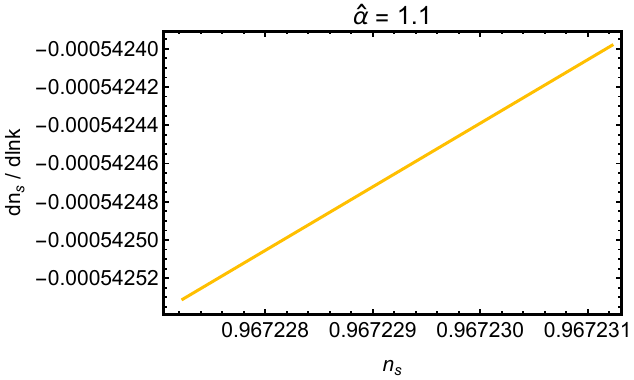}
\caption{\label{fig3} Inflationary predictions of the potential in \eqref{canonicpot} for $\hat{\alpha} = 1.1$. Top panel shows $c_O-n_s$ (left) and $c_O-r$ (right). Bottom panel shows $n_s-(\mathrm{d}n_s/\mathrm{d}\ln k)$.}
\end{figure}

Furthermore, figures \ref{fig3} and \ref{fig5} show that in order to produce inflationary predictions which are consistent with the current data, the quadratic curvature coefficient $c_O$ values should be in the range $-1.020 \times 10^{-10} \lesssim c_O \lesssim -1.0165 \times 10^{-10}$ for $\hat{\alpha} = 1.1$ and $-1.03576 \times 10^{-10} \lesssim c_O \lesssim -1.03572 \times 10^{-10}$ for $\hat{\alpha} = 0.95$. 

It is important to note that, there is no solution to provide inflationary mechanism for $ c_O > 0$ values. This is noteworthy in that, the analysis excludes positive $c_O$ values which result in scalaron (inflaton) as a tachyon.   Note also that, since $c_O$ is a loop induced parameter of symmmergent gravity which depends on $(n_b-n_f)$, excluding the positive values of $c_O$ for the successful inflation means a (SM+BSM) matter sector wherein fermions outnumber the bosons.

In addition, the plots present that for $\hat{\alpha} > 1$ cases, we have reasonable results of inflationary predictions only for the values $1.01 \lesssim \hat{\alpha} \lesssim 1.15$. It is important to mention that for $\hat{\alpha} \gtrsim 1.15$ values, there is no solution to satisfy the inflationary predictions for our model. However, there are no such limits to have a satisfactory inflationary parameters for $\hat{\alpha} < 1$ values. 

Finally, we show the results of $\mathrm{d}n_s/\mathrm{d}\ln k$ for our model in the bottom panels of figures \ref{fig3} and \ref{fig5}. Current observational resolutions for the running of the spectral index are not good enough to definitely rule out a model but with the improvements of future observations, such as 21 cm line \cite{Kohri:2013mxa,Basse:2014qqa,Munoz:2016owz}, we expect strong constraints to be able to test the potential model.


In summary, the numerical results show that our model has fermion surplus because we have only solutions for $ c_O < 0$ values. Also, for these $c_O$ values, we display our results for the effect of bosons and fermions dominance in terms of masses on the inflationary predictions. It can be emphasized that our model has an excess of fermion number, also the predictions are in good agreement with the latest constraints for both fermions' and bosons' mass dominance. For our model, we find that $n_s-r$ values are very close to the results of $R^2$ inflation model and both results agree very well with the current data.

\begin{figure}\label{voless1CL}
\centering
\includegraphics[width=.36\textwidth]{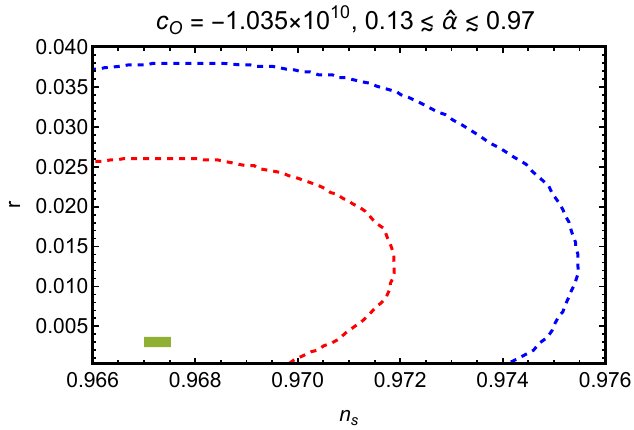} \ \
\includegraphics[width=0.46\textwidth]{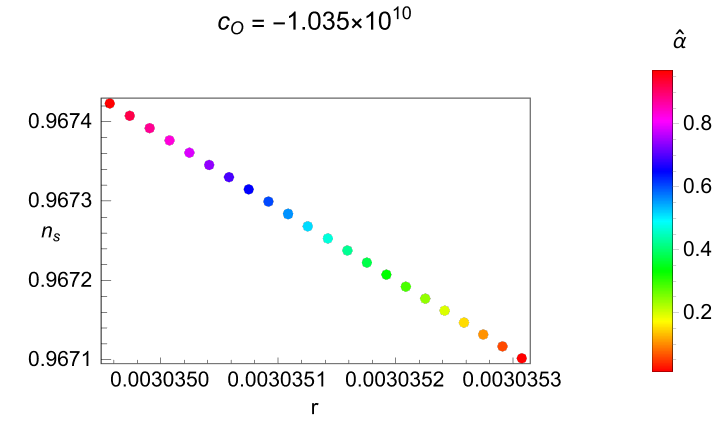}
\caption{\label{fig4}   Inflationary predictions of the potential in \eqref{canonicpot}. The left panel shows $n_s-r$ values for $c_O = -1.035 \times 10^{10}$ and $0.13 \lesssim \hat{\alpha} \lesssim 0.97$ and blue(red) contours indicate the recent 95\%(68\%) CL given by BICEP/Keck \cite{BICEP:2021xfz}. Right panel shows $r-n_s$ values, where each dot corresponds to a different $\hat{\alpha}$ value (color coded) but to the same  $c_O = -1.035 \times 10^{10}$.}
\end{figure}

\begin{figure}\label{voless1nsr}
\centering
\includegraphics[width=.79\textwidth]{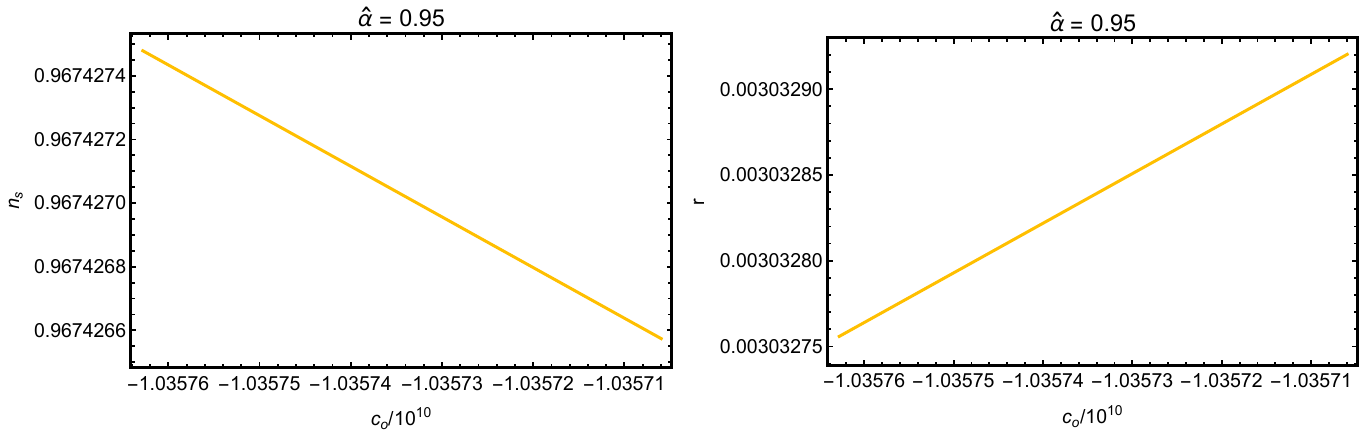}\\
\includegraphics[width=.42\textwidth]{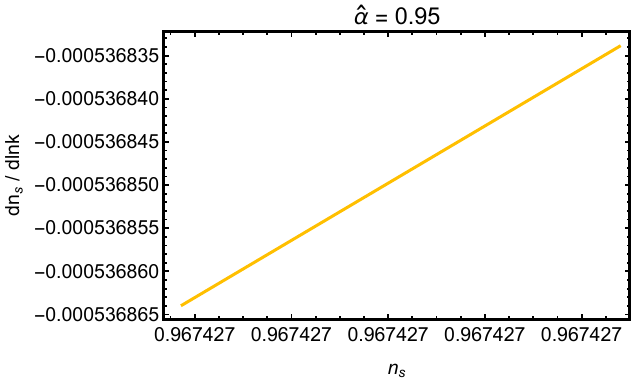}
\caption{\label{fig5} Inflationary predictions of the potential in \eqref{canonicpot} for $\hat{\alpha} = 0.95$. Top panel shows $c_O-n_s$ (left) and $c_O-r$ (right). Bottom panel shows ($n_s-\mathrm{d}n_s/\mathrm{d}\ln k)$.}
\end{figure}



\section{Conclusion}

\label{conc}

Symmergent gravity is a new emergent gravity framework that merges SM and gravity and concurrently restores the broken gauge symmetries of the QFT. In this paper, we have studied the cosmological inflation in symmergent gravity because cosmological phenomena are legitimate test-beds for gravitational theories. We have constrained the symmergent gravity parameters $c_0$ and $\hat{\alpha}$, whose origins are the loop-induced interactions of matter sector in flat spacetime. The loop factor $c_0$ related to the quartic corrections ($\Lambda^4 $) turns into the coefficient of quadratic curvature term in symmergent gravity. This coefficient carries information about the difference of total fermion and boson numbers ($n_b-n_f$) in the underlying QFT of SM together with the BSM sector. The parameter $\hat{\alpha}$ in vacuum energy is a measure of deviation from the characteristic mass scale of the QFT for the boson and fermion masses.  

According to our analysis, to be able to satisfy the inflationary mechanism $c_0$ should be a negative number at the order of $10^{10}$ and this corresponds to an inflationary epoch where fermions outnumber bosons in the matter sector. In the first place, this is not startling at all from a theoretical standpoint because a theory which excludes scalaron as a tachyon should have a negative quadratic curvature coefficient if the theory is written using the West-coast metric signature. Second, as an effective theory, SM already  has fermion number dominance. Moreover, recent black hole studies \cite{Rayimbaev:2022hca} report parallel results. Furthermore, our analysis shows that both cases of fermion mass dominance ($\hat{\alpha}<1$) and boson mass dominance ($\hat{\alpha}>1$) satisfy the inflationary period as long as the deviation from the mass degeneracy in the matter sector is minute.

As a result, the analysis is novel in that $c_O$ is proportional to $n_b-n_f$ in the underlying QFT and enables therefore probing of the QFT spectrum via the inflationary dynamics. The novelty comes also from the vacuum energy $V_0$, which indicates the deviation from mass-degeneracy with $\hat{\alpha}$. It is shown that for successful inflationary period it does not matter whether the matter sector has fermion mass dominance or boson mass dominance.



\acknowledgments
The authors thank Durmu\c{s} Demir for useful discussions on the symmergent gravity theory. The work of O. S. is supported by the T{\"U}B{\.I}TAK B{\.I}DEB-2218 national postdoctoral fellowship grant 118C522.





\end{document}